\documentclass[letterpaper,10pt,conference]{ieeeconf}
\IEEEoverridecommandlockouts
\usepackage{cite}
\usepackage{amsmath,amssymb,amsfonts}
\usepackage{algorithmic}
\usepackage[hyphens]{url}\usepackage{rotating} 
\usepackage{graphicx}
\usepackage{textcomp}
\usepackage{xcolor}
\usepackage{multirow}
\usepackage{tikz}
\newcommand{\gray}[1]{\textcolor{gray}{#1}}
\def\BibTeX{{\rm B\kern-.05em{\sc i\kern-.025em b}\kern-.08em
    T\kern-.1667em\lower.7ex\hbox{E}\kern-.125emX}}
\begin{document}

\title{\LARGE\bfseries
Towards Mode-Balanced Intersection Control:
Hierarchical Coordination of Signal Timings,
AV Trajectories, and Cyclist Speed Advice
}

\author{Wies Siffels, Giray Önür, Azita Dabiri%
\thanks{The authors are with the Delft Center for Systems
and Control, Delft University of Technology, Delft,
The Netherlands
{\ttfamily\small wiessiffels@hotmail.nl},
{\ttfamily\small \{g.oenuer, a.dabiri\}@tudelft.nl}.}
}

\maketitle

\begin{abstract}
Connected and automated vehicles (AVs) enable traffic signals and vehicle trajectories to be optimised jointly, but existing co-optimisation frameworks overlook cyclists, even though they share the same intersections. We propose a hierarchical framework that optimises traffic signal timings, AV trajectories, and cyclist speed advice to reduce travel times for both motorised vehicles and cyclists. The upper level jointly optimises signal timings and cyclist speed advice, while the lower level optimises AV trajectories and also updates cyclist speed advice obtained from the higher level at a higher frequency based on the resulting signal plan. The framework is evaluated in simulation under multiple demand profiles and optimisation parameter settings. Compared with optimized vehicle trajectories with fixed-time signal control, the proposed framework reduces mean travel times by 13.6\% for AVs and 12.3\% for cyclists. Adding cyclist speed advice further reduces mean cyclist waiting time by about 80\% relative to the same framework without speed advice. These results demonstrate substantial benefits for both modes of transport.
\end{abstract}

\section{Introduction}
\label{chap:intro}

Cycling is an important mode of urban transport, offering a sustainable, healthy, and space-efficient alternative to car travel \cite{pucher2008making}.
The Netherlands is a prominent example, where about 27\% of all trips were made by bike in 2023 \cite{bike_stats}.
In Dutch cities, signalised intersections are therefore shared by large flows of both cyclists and motorised vehicles, whose conflicting movements compete for the same green time.
Allocating green time is inherently a trade-off: extending the green phase for one mode increases waiting times for the other.
Traffic signal control has traditionally been designed around motorised traffic, while waiting at red lights is a major source of delay and discomfort for cyclists \cite{fioreze2019perceived, broach2012cyclists}. Although some previous studies have incorporated cyclists into intersection control frameworks \cite{wang2023human, arora2025mathematical, portilla2016model, niels2020integrated}, research on signal optimisation that explicitly accounts for cyclists remains limited .
In cities that aim to promote cycling, signal control should therefore consider the performance of both modes rather than prioritise one at the expense of the other.

The introduction of connected and automated vehicles (AVs) adds a new dimension to this trade-off.
Unlike conventional vehicles, AVs can follow trajectories that are optimised jointly with signal timings, allowing motorised traffic to pass through intersections more efficiently.
Signal optimisation and AV trajectory optimisation have been extensively studied in full AV environments \cite{li2014signal, wang2023coupled, li2025traffic, dabiri2025optimization}, and more recent work has addressed mixed traffic comprising AVs and regular vehicles (RVs), with RVs modelled as non-controllable agents \cite{li2024managing, pourmehrab2019optimizing, zou2024integrated}.
However, these frameworks focus on motorised traffic.
If such efficiency gains are pursued without considering cyclists, the coordination of signals and AVs risks shifting the balance at intersections further towards motorised traffic.
Although full AV penetration is not expected in the near term, a full AV environment represents the setting in which the efficiency potential of vehicle--signal coordination is greatest, and hence the setting in which the risk of disadvantaging cyclists is most pronounced.
Examining this setting therefore indicates whether cyclists can still benefit when motorised traffic is fully controllable, and that's one of the goals of this paper.

Moreover, although cyclists cannot be controlled in the same way as AVs, they can be supported in adapting to the signal plan.
Speed advice (SA) helps cyclists adjust their arrival times to upcoming green intervals.
This concept has been  studied for motorised vehicles through Green Light Optimal Speed Advisory (GLOSA) \cite{katsaros2011performance}, and more recent work has applied it to cyclists \cite{ dabiri2019automatic, dabiri2020optimized, dabiri2019optimal}.
However, these cyclist SA approaches do not consider the optimisation of signal timings and AV trajectories.
A gap therefore remains in coordinating signal timings, AV trajectories, and cyclist SA within a common framework that serves both modes.

To address this gap, this paper proposes a hierarchical optimisation framework in which the upper level jointly optimises signal timings and cyclist speed advice, while the lower level updates cyclist SA and optimises AV trajectories at a higher frequency.
The framework aims to improve the traffic performance of both AVs and cyclists, rather than improving one mode at the expense of the other.

The remainder of this paper is organised as follows.
Section \ref{chap:problem} formulates the traffic control problem.
Section \ref{chap:methodology} describes the proposed hierarchical optimisation framework.
Section \ref{chap:results} presents the experiments and discusses the results.
Finally, Section \ref{chap:conclusion} concludes the paper.

\section{Problem formulation} \label{chap:problem}
In this section, the traffic control problem is formulated for a general intersection scenario, whereas the specific details are described in Section \ref{chap:results}. The following assumptions are used in formulating this scenario:

\begin{enumerate}
    \item All vehicles are fully autonomous, communicate their positions and velocities to each other and the infrastructure, and accurately control their accelerations.
    \item Approaching the intersection, all vehicles move on a predetermined path that does not require them to switch lanes.
    \item Road users leaving the intersection are not constrained by downstream traffic.
    \item All road users obey the traffic signals and the speed advice signs, unless stated otherwise.
\end{enumerate}

\begin{figure}[htb]
    \centering
    \input{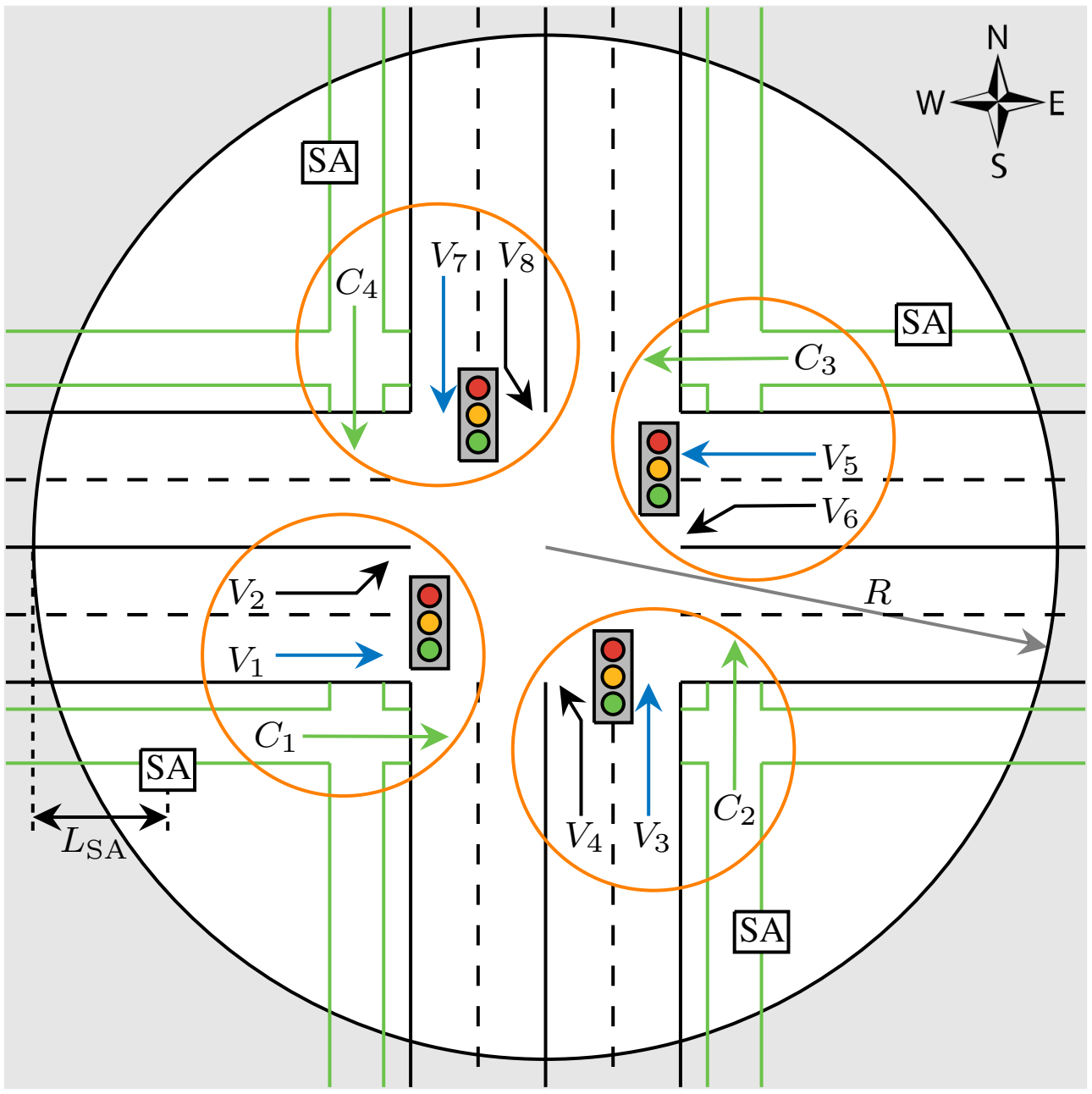}    
    \caption{Example of an intersection layout. The orange circles indicate a possible grouping of movements into signal phases. SA indicates speed advice}
    \label{fig:intersection}
\end{figure}

To make the signal timing (ST) adaptable to changing conditions, the cycle length is flexible. There is assumed to be a fixed set of signal phases, $\mathcal{P} = \{1, \ldots, \Phi\}$, where $\Phi$ is the total number of phases. Each signal phase has a set of non-conflicting movements which are allowed to cross the intersection when this phase is green.  Each movement is assigned to one signal phase.  To further ensure safety, amber time is included for cyclists, and the signal phases are separated by a fixed clearance time, during which all phases are red. The traffic controller determines the green times, which are bounded by minimum and maximum values. A possible layout of a four-legged intersection\footnote{A four-legged intersection is used here to illustrate the problem formulation. The proposed framework is not restricted to this layout.} is shown in Figure \ref{fig:intersection}.

Vehicles are detected when they enter the communication range $R$, which is centred around the intersection. All vehicles communicate their positions and velocities with an intersection control unit. The intersection control unit is assumed to be able to communicate acceleration commands to autonomous vehicles within the communication range. Since the vehicles do not need to change lanes or overtake (see Assumption 2), their accelerations are calculated via the intersection controller, assuming the vehicles move longitudinally but not laterally on the road. There are a total of $M^\mathrm{V}$ vehicle movements, where the superscript 
$V$ denotes an index identifying vehicle movements. These movements belong to the set $\mathcal{M^\mathrm{V}}= \{1, \ldots , M^\mathrm{V}\}$. 

A separate lane for cyclists is located on the right side of the road. There are $M^\mathrm{C}$ cyclist movements, where the superscript 
$C$ denotes an index identifying cyclist movements. These movements belong to the set $\mathcal{M^\mathrm{C}}= \{1, \dots, M^\mathrm{C} \}$. Cyclists are assumed to not affect each other's movements and are free to overtake. The infrastructure detects the cyclists when they enter the communication range and measures their positions and velocities.

\section{Solution approach} \label{chap:methodology}
\begin{figure}[htb]
    \centering
    \includegraphics[width=\linewidth]{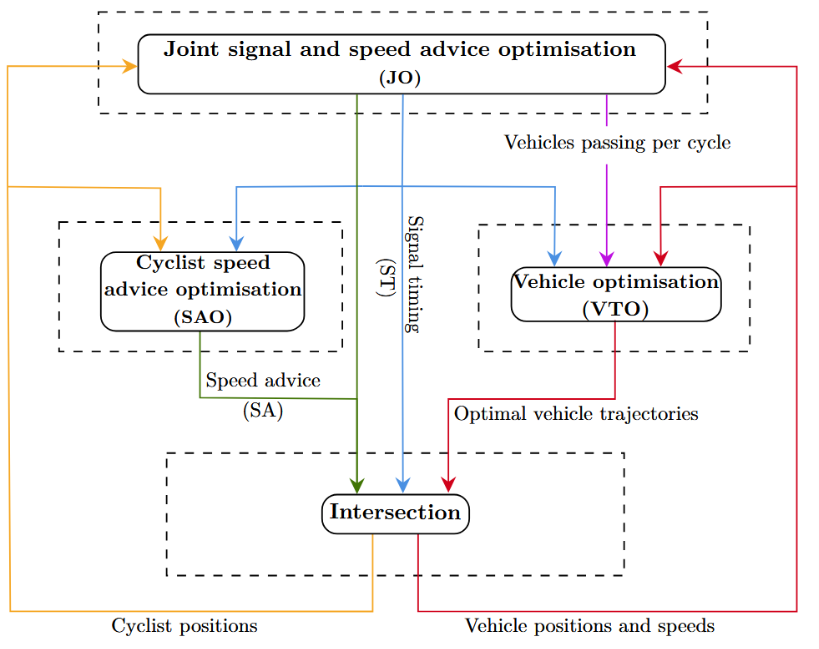}
    \caption{Hierarchical optimisation structure.}
    \label{fig:Hier_structure}
\end{figure}

The hierarchical optimisation consists of three components: i) the joint optimisation of speed advice and signal timing (JO), ii) the speed advice optimisation (SAO), iii) and the vehicle trajectory optimisation (VTO) . The hierarchical framework is depicted in Figure \ref{fig:Hier_structure}. The JO is performed on the upper level at a lower frequency and finds the optimal signal timing, the optimal number of vehicles to pass during the green times, and the speed advice for cyclists. The current positions of cyclists and the positions and velocities of vehicles are used as inputs to this block. The SAO updates the speed advice at a higher frequency and is executed only when the JO is not, as the JO already incorporates the optimisation of the SA. This allows the advice to be tailored more frequently to smaller groups of cyclists, enabling more targeted guidance.
The vehicle optimisation block (VTO) computes the optimal vehicle trajectories based on the current positions and velocities of the vehicles, as well as the optimal signal timing and the number of vehicles to pass during the green phases, as determined by the JO. 

The speed advice update frequency, $f_\mathrm{SA}$, is chosen to be higher than the signal update frequency, $f_\mathrm{sig}$, considering that traffic conditions change more slowly than cyclist dynamics. Moreover, the speed advice should not change too frequently, since cyclists need time to perceive the advice and adapt to it, and the speed advice update frequency is therefore lower than the vehicle update frequency, $f_\mathrm{veh}$. Based on these considerations, we use the following control frequency hierarchy:
\begin{equation}
    f_\mathrm{sig} < f_\mathrm{SA} < f_\mathrm{veh}
\end{equation}

\subsection{Joint optimisation of speed advice and signal timing (JO)}
The joint optimisation of speed advice and signal timing is formulated as a mixed-integer linear program (MILP) with decision variables $\mathbf{G}_\mathrm{sig}$ (vector of green durations in the horizon), $\mathbf{X}_\mathrm{veh}$ that collects vehicle passing times and binary cycle assignments, $\mathbf{X}_\mathrm{cyc}$ that collects cyclist waiting times and binary cycle assignments, and $\mathbf{S}$ indicating speed advice slowness factors. The binary cycle assignment variables in $\mathbf{X}_\mathrm{veh}$ and $\mathbf{X}_\mathrm{cyc}$ determine the signal cycle during which each vehicle or cyclist is assigned to cross the intersection. The slowness factors in $\mathbf{S}$ determine the advised cycling speeds, allowing cyclists' arrival times at the intersection to be adjusted to better align with the signal timing.

There are three categories of constraints to be considered: constraints related to signal timing, constraints associated with vehicles, and those associated to cyclists.\\

\noindent Signal timing should respect the following constraints where $l\in\mathbb{I}_{[1,L]}$ indicates the cycle:
\begin{gather}
    \label{eq:g_min_max}
   g_\mathrm{min}\le g_{\phi,l} \le g_\mathrm{max} ,\quad \forall \phi \in \mathcal{P}\\
    \label{eq:t_clr}
    t_{\phi+1,l}^\mathrm{sg} -  t_{\phi,l}^\mathrm{eg}  = t_\mathrm{amb} + t_\mathrm{clr}, \quad \forall \phi \in \mathbb{I}_{[1,\Phi-1]}\\
    \label{eq:t_clr2}
     t_{1,l+1}^\mathrm{sg} -t_{\Phi,l}^\mathrm{eg} = t_\mathrm{amb} + t_\mathrm{clr}
\end{gather}
Constraints \eqref{eq:g_min_max}\;--\;\eqref{eq:t_clr2} incorporate the bounds for the green durations of cycle $l$ in phase $\phi$ denoted by $g_{\phi,l}$, and the amber and all-red intervals between consecutive green phases, where $t^\mathrm{sg}_{\phi,l}$ and $t^\mathrm{eg}_{\phi,l}$ are the start time and end time of the green phase $\phi$ in cycle $l$. These depend on the green durations through $t^\mathrm{eg}_{\phi,l} - t^\mathrm{sg}_{\phi,l} = g_{\phi,l}$. 
\noindent\\
Vehicle $V(i)$ for $i\in\mathbb{I}_{[1,N_m^{\mathrm{V}}]}$ should satisfy the following constraints ($m\in\mathcal{M}^{\mathrm{V}}$, unless stated otherwise):
\begin{gather}
\label{eq:tau_min_constraint}
 \tau^{\mathrm{V}(i)}_m \ge \tau^{\mathrm{V}(i)}_{\mathrm{min}, m} + g_\mathrm{elap}\\
\label{eq:headway}
     \tau_{m}^{\mathrm{V}(i)} - \tau_m^{\mathrm{V}(i-1)} \ge t_\mathrm{hw}, \quad \forall i \in \mathbb{I}_{[2, N_m^\mathrm{V}]}\\
     \label{eq:veh_green_constraint}
    \tau^{\mathrm{V}(i)}_m \in G_{\phi^\mathrm{V}(m)}\\
    \label{eq:sum_b}
    \sum_{l=1}^{L-1} b^{\mathrm{V}(i)}_{m,l} \le 1 \\
    \label{eq:veh_safety}
    b^{\mathrm{V}(i)}_{m,1} \geq \tilde{b}_m^{\mathrm{V}(i)}, \quad \forall m \in \mathcal{M}^{\mathrm{V}} : \phi^\mathrm{V}(m)=\phi_\mathrm{curr}
\end{gather}
 where $\phi_\mathrm{curr}$ denotes the currently green phase and the mapping $\phi^\mathrm{V}(m)$ gives the signal phase assigned to vehicle movement $m$.
 The vehicle arrival times, $\tau^{\mathrm{V}(i)}_m $, are constrained by a minimum arrival time, $\tau^{\mathrm{V}(i)}_{\mathrm{min}, m}$, in \eqref{eq:tau_min_constraint} as defined in \cite{xu2018cooperative}\footnote{\cite{xu2018cooperative} computed the minimum arrival time with the current position and velocity of the vehicle.}. The time elapsed in the current green phase at the instant of optimisation, $g_\mathrm{elap}$, is added to the minimum arrival time so that arrival times are measured from the start of the current green phase. For safety reasons, all vehicles must maintain a certain headway from the vehicle in front of them. This is incorporated into \eqref{eq:headway}, by maintaining the headway in time, $t_\mathrm{hw}$, between the arrival times. Note that the safety constraint over the entire trajectory will be imposed in the lower level. Constraint \eqref{eq:veh_green_constraint} ensures all vehicles only pass during green, where $G_{\phi} = \bigcup_{l=1}^{L}[t_{\phi,l}^\mathrm{sg}, t_{\phi,l}^\mathrm{eg}]$ is a union set of all the green phases over $L$ cycles. The green-interval constraints \eqref{eq:veh_green_constraint} use big-$M$ formulations via variable $b_{m,l}^{\mathrm{V}(i)}$ indicating the set membership: $b_{m,l}^{\mathrm{V}(i)}=1$ enforces crossing in cycle $l$, while setting all $L-1$ binaries to zero selects cycle $L$ (Equation \eqref{eq:sum_b}). We also introduce $\tilde{b}$ to account for the situation where a vehicle is unable to stop for a red light in the current green phase. $\tilde{b}$ is an input to the optimization problem and equal to 1 when the minimum stopping distance of the vehicle (after maximum brake) is beyond the stop line. Then  \eqref{eq:veh_safety} enforces the passing of that vehicle in the current green phase.\\ 
\noindent
For cyclists, we first need to model the arrival time of a cyclist to the intersection. We assume that the cyclist will fully comply with the speed advice (the effects of
relaxing this assumption are evaluated in the case study). Given this assumption, the arrival time of cyclists that have passed the SA sign is computed by  \eqref{eq:t_TL}:
\begin{equation}
    t_{\mathrm{arr},m}^{\mathrm{TS}(j)} = \frac{R - p^{\mathrm{C}(j)}_{m}}{v^{\mathrm{C}(j)}_{\mathrm{SA},m}} + g_\mathrm{elap}, \; \forall  j \in \mathbb{I}_{[1, o_m]}\label{eq:t_TL}
\end{equation}
where $o_m$ denotes the number of cyclists in movement $m$ that have already passed the SA sign.
For the cyclists that have not yet received SA, the time to get to the SA sign is determined by \eqref{eq:t_SA}. This is then used to determine the arrival time at the signals in \eqref{eq:t_arr_with_adj}.
\begin{gather}
    \label{eq:t_SA}
    t_{\mathrm{arr},m}^{\mathrm{SA}(j)} = \frac{L_\mathrm{SA} - p^{\mathrm{C}(j)}_{m}}{v^\mathrm{C}_\mathrm{nor}}, \; \forall  j \in \mathbb{I}_{[o_m+1,N^\mathrm{C}_m]}\\
       t^{\mathrm{C}(j)}_{\mathrm{arr},m} =  \; t_{\mathrm{arr},m}^{\mathrm{SA}(j)} + S^r_m \frac{R - L_\mathrm{SA} }{ v^\mathrm{C}_\mathrm{nor}} + 0.5(1-S^r_m)t_\mathrm{adj}+g_\mathrm{elap}, \label{eq:t_arr_with_adj}      
\end{gather}
with 
\begin{equation}
     t_{\mathrm{arr},m}^{\mathrm{SA}(j)} \in \left[\frac{r-1}{f_\mathrm{SA}},\frac{r}{f_\mathrm{SA}}\right],\quad j\in\mathbb{I}_{[o_m+1,N_m^\mathrm{C}]}
     \label{eq:bound-tSA}
\end{equation}     
In \eqref{eq:t_arr_with_adj} the slowness factor $S_m^r=v_\mathrm{nor}^\mathrm{C}/v_\mathrm{SA}^\mathrm{C}$ is the ratio of normal to advised cycling speed, where $S_m^r=1$ represents normal speed. A horizon of $N_\mathrm{adv}$ SAs is optimised, taking into account all cyclists within the communication range, $R$. The SA sign is located at a distance of $L_\mathrm{SA}$ from the point at which cyclists enter the communication range. Which SA is received by a cyclist is dependent on the arrival time at the SA sign and is determined by \eqref{eq:bound-tSA}, where the $r^\mathrm{th}$ SA is received. The cyclists adjust their speed linearly towards the advised speed over an adjustment period, $t_\mathrm{adj}$. It keeps the cyclist constraints linear and is bounded by:
\begin{equation}
\label{eq:S_bounds}
     S_\mathrm{min} \le S^r_m \le S_\mathrm{max}, \quad \forall r \in \mathbb{I}_{[1,N_\mathrm{adv}]}
\end{equation}
\noindent
Constraint \eqref{eq:cyc_green} then ensures that cyclists cross only during green or the permitted part of amber, where $G^\mathrm{C}_{\phi} = \bigcup_{l=1}^{L}[t_{\phi,l}^\mathrm{sg}, t_{\phi,l}^\mathrm{eg} + t_\mathrm{amb}-1]$ is the union set of the permitted crossing intervals. For cyclists, the amber time, $t_\mathrm{amb}$, is considered to ensure that they can stop for a red light. 
\begin{gather}
      \tau_{\mathrm{pass},m}^{\mathrm{C}(j)} = \tau^{\mathrm{C}(j)}_{\mathrm{arr,m}} + \tau^{\mathrm{C}(j)}_{\mathrm{wait},m}   \in G^\mathrm{C}_{\phi^\mathrm{C}(m)}  \nonumber\\*
      \label{eq:cyc_green}
      \mathrm{with} \quad \mathbf{\tau}^{\mathrm{C}(j)}_{\mathrm{arr,m}} = 
    \begin{cases}
         t_{\mathrm{arr},m}^{\mathrm{TS(j)}}, \quad \mathrm{if}\; j \le o_m   \\
         t^{\mathrm{C}(j)}_{\mathrm{arr},m} , \quad \mathrm{if}\; j > o_m
    \end{cases} 
\end{gather}
where $\phi^\mathrm{C}(m)$ gives the signal phase assigned to cyclist movement $m$. Similarly to \eqref{eq:sum_b}, \eqref{eq:sum_b_cyc} ensures that cyclists pass the signals only once. Moreover,  \eqref{eq:wait_min} ensures that the waiting time for cyclists is not negative:    
\begin{gather}
          \label{eq:sum_b_cyc}
    \sum_{l=1}^{L-1} b^{\mathrm{C}(j)}_{m,l} \le 1\\
    \label{eq:wait_min}
    \tau^{\mathrm{C}(j)}_{\mathrm{wait},m} \ge 0 
\end{gather}
Hence, the optimization problem of the JO level can be written as 
\begin{equation}
\label{eq:joint_opt}
\begin{aligned}
&\{\mathbf{G^*_\mathrm{sig}}, \mathbf{X}_\mathrm{veh}^*, \mathbf{X}_\mathrm{cyc}^*, \mathbf{S^*}\} =\\
&\operatorname*{argmin}\; W^\mathrm{V}\sum_{m=1}^{M^\mathrm{V}}\sum_{i=1}^{N_m^\mathrm{V}}\tau_m^{\mathrm{V}(i)}
+W^\mathrm{C}\sum_{m=1}^{M^\mathrm{C}}\sum_{j=1}^{N_m^\mathrm{C}}\tau_{\mathrm{wait},m}^{\mathrm{C}(j)}\\
& \text{subject to: } \eqref{eq:g_min_max}-\eqref{eq:wait_min}
\end{aligned}
\end{equation}
where $W^\mathrm{V}$ and $W^\mathrm{C}$ are the weights on total AV arrival time and total cyclist waiting time, respectively.
\subsection{Speed advice optimisation (SAO)}
Between successive JO executions, with ST fixed, SAO optimises only the cyclist variables and speed advice, subject to the cyclist constraints:
\begin{align*}
\{ \mathbf{X}_\mathrm{cyc}^* , \mathbf{S}^* \}  &=   \mathrm{argmin}\sum_{m=1}^{M^\mathrm{C}}\sum_{j=1}^{N_m^\mathrm{C}}\tau_{\mathrm{wait},m}^{\mathrm{C}(j)},\\
& \text{subject to: } \eqref{eq:t_TL}-\eqref{eq:wait_min}
\end{align*}

\subsection{Vehicle trajectory optimisation (VTO)}
 As inputs, in addition to the initial vehicle states, the vehicle trajectory optimisation uses the signal timing and the number of vehicles passing during the green intervals for each movement, both determined by the JO. Given this input, it then determines vehicle trajectories by optimising accelerations subject to signal timing and safety constraints. Since the vehicle movements are grouped in non-conflicting phases, the vehicle optimisation can be performed independently for each movement, allowing the vehicle trajectories of different movements to be computed in parallel. 

The optimisation is formulated as a discrete-time problem with a prediction horizon time of $t_\mathrm{f} = H \Delta t$. The dynamical model of a vehicle is described by
\begin{align}
\label{eq:dyn_model_veh}
    &\mathbf{x}_{m,k+1}^{(i)} = F^{(i)}_{m}(\mathbf{x}_{m,k}^{(i)}, u_{m,k}^{(i)}, \Delta t), &k \in \mathbb{I}_{[0, H-1]}\\
    &c_{m}^{(i)}(\mathbf{x}_{m,k}^{(i)}, u_{m,k}^{(i)}) \le 0, &k \in \mathbb{I}_{[0, H-1]} 
\label{eq:dyn_model_veh2}
\end{align}
where the position of the $i$th vehicle in movement $m$ at time step $k$, $p_{m,k}^{\mathrm{V}(i)}$, and its velocity, $v_{m,k}^{\mathrm{V}(i)} $, are collected in the state vector $\mathbf{x}_{m,k}^{(i)} = [p_{m,k}^{\mathrm{V}(i)}, v_{m,k}^{\mathrm{V}(i)}  ]^\top$. The state and input constraints are represented by the function $c_{m}^{(i)}$. The input, $u_{m,k}^{(i)}$, is the acceleration. To avoid rear-end collisions, a certain headway should be maintained between two vehicles. This spatial headway, $d_\mathrm{hw}$, is formulated in the following inequality constraint:
\begin{equation}
\label{eq:veh_hw}
    p_{m,k}^{\mathrm{V}(i+1)} +d_\mathrm{hw} \le p_{m,k}^{\mathrm{V}(i)} 
\end{equation}
The states and inputs are collected in $\mathbf{w}_{m,k}^{(i)} = (\mathbf{x}_{m,k}^{(i)}, u_{m,k}^{(i)})$. The complete state and input trajectory is denoted by $\mathbf{w}_{m}^{(i)}$. Vehicles are only allowed to cross the intersection during green. The notation $p_{m}^{(i)}(t, \mathbf{w}_{m}^{(i)})$ denotes the position of vehicle $i$ in movement $m$ at time $t$, given the inputs $\mathbf{w}_{m}^{(i)}$. These constraints are formulated as:
\begin{align}
\label{eq:green1}
    p_{m}^{(i)}(t_{\phi^\mathrm{V}(m),l}^\mathrm{sg}, \mathbf{w}_{m}^{(i)}) \le R,  &\quad  \forall i \in \mathbb{I}_{[B_{m,l-1}+1,N^\mathrm{V}_m]}\; \; \; \; \;\\
    \label{eq:green2}
     p_{m}^{(i)}(t_{\phi^\mathrm{V}(m),l}^\mathrm{eg}, \mathbf{w}_{m}^{(i)}) \ge R,& \quad    \forall i \in \mathbb{I}_{[1,B_{m,l}]}\; \; \; \; \; \; \;\\
    \label{eq:green3}
    p_{m}^{(i)}(t_{\phi^\mathrm{V}(m),l+1}^\mathrm{sg}, \mathbf{w}_{m}^{(i)}) \le R,  &\quad \forall i \in \mathbb{I}_{[B_{m,l}+1,N^\mathrm{V}_m]}
\end{align}
where $B_{m,l}$ denotes the total number of vehicles in movement $m$ assigned by the JO to pass the stop line from the beginning of the prediction horizon through the end of cycle $l$. Constraint \eqref{eq:green1} ensures that vehicles that are not assigned to earlier cycles are still behind the stop line at the start of the green (to avoid red light crossing). Constraint \eqref{eq:green2} ensures that all $B_{m,l}$ vehicles have passed the stop line at the end of the green. Finally, \eqref{eq:green3} ensures that all remaining vehicles do not pass the stop line until the start of the next green. The following objective function\footnote{Note that the trajectories can be optimised for different performance metrics from those used in the joint optimisation, such as fuel consumption.} is used in the VTO block:
\begin{equation}
\label{eq:obj_func_veh}
    o_{m}^{(i)}(\mathbf{w}_{m}^{(i)}) = \sum^{H-1}_{k=0} \left( Q^\mathrm{u} (u_{m,k}^{(i)})^2 + Q^\mathrm{v} (v_{m,k}^{(i)} - v^\mathrm{d})^2 \right),
\end{equation}
where the term $Q^\mathrm{u} (u_{m,k}^{(i)})^2$ penalises large acceleration inputs, while the term $Q^\mathrm{v} (v_{m,k}^{(i)} - v^\mathrm{d})^2$ penalises deviations from the desired speed, $v^\mathrm{d}$. The state and input trajectories, $\mathbf{w}_{m}^{(i)}$, are combined in the vector $\mathbf{W}_m$. The resulting nonlinear vehicle trajectory optimisation problem for a single movement $m$ is formulated as follows:
\begin{equation*}
\begin{aligned}
&\mathbf{W}_m^* = \operatorname*{argmin}\sum_{i=1}^{N^\mathrm{V}_m}o_m^{(i)}(\mathbf{w}_m^{(i)})\\
&\text{subject to: } \eqref{eq:dyn_model_veh},\eqref{eq:dyn_model_veh2},\quad \forall i\in\mathbb{I}_{[1,N_m^\mathrm{V}]}\\
&\phantom{\text{subject to: }}\eqref{eq:veh_hw},\eqref{eq:green1},\eqref{eq:green2},\eqref{eq:green3}.
\end{aligned}
\end{equation*}

\section{Results} \label{chap:results}
The performance of the proposed hierarchical framework is evaluated in simulations in terms of the waiting time for cyclists and the travel time for both cyclists and vehicles. The cyclist dynamics used for the simulations are discussed in Section \ref{subsec:sim_cyc}. The case study used to assess the performance is explained in Section \ref{subsec:case}. In Section \ref{subsec:weight}, the impact of the weight ratio is assessed, and the performance of the proposed framework is compared across different simulation durations. Finally, the impact of the speed advice on the performance is discussed in Section \ref{subsec:res_SA}.

\subsection{Simulation of cyclist dynamics} \label{subsec:sim_cyc}

Cyclist trajectories are simulated using a discrete-time kinematic model consisting of three dynamic modes. The active mode depends on the cyclist's position, the speed advice and the traffic signal. In the first mode, cyclists maintain their current speed. In the second mode, cyclists adjust their speed to follow the speed advice or accelerate after coming to a stop. The acceleration in this mode is determined by

\begin{equation} \label{eq:adj_speed} u_\mathrm{adj}(t) = u_\mathrm{des}\left(1-\left(\frac{v^c(t)}{v_\mathrm{SA}} \right)^2\right), \end{equation}

\noindent where $v^c(t)$ denotes the cyclist's speed at time $t$, $v_\mathrm{SA}$ is the advised speed and $u_\mathrm{des}$ is a parameter determining the desired acceleration. When the advised speed is lower than the current speed, cyclists are assumed to reduce their speed by rolling out rather than braking. Therefore, the deceleration is limited to $u_\mathrm{max,roll}$.

In the third mode, cyclists decelerate to a full stop when approaching a red signal. The required deceleration is determined by
\begin{equation}\label{eq:cyc_stop} u_\mathrm{red}(t) = -\frac{v^c(t)^2}{2d}, \end{equation}

\noindent where $d$ denotes the remaining distance to the stop line. The deceleration is limited by the maximum braking capability of the cyclist, $u_\mathrm{max,decel}$.

The acceleration determined by the active dynamic mode is used to update the cyclist's speed and position at each simulation time step, $\Delta$t. The speed and position are updated according to the discrete-time equations of motion:
\begin{align}
v^c(t+\Delta t) &= v^c(t) + u(t)\Delta t, \\
p^c(t+\Delta t) &= p^c(t) + v^c(t)\Delta t + \frac{1}{2}u(t)\Delta t^2, \end{align}

\noindent where $u(t)$ represents the acceleration determined by the active dynamic mode. The model assumes that cyclists do not cross the stop line during a red signal. The spatial regions associated with the three dynamic modes are illustrated in Figure~\ref{fig:cyc_regions} in the Appendix.

\subsection{Case study}
\label{subsec:case}
The intersection in Figure \ref{fig:intersection} is considered in the case study. The performance of the proposed hierarchical framework is evaluated under three demand profiles. The demand profiles are given in Table \ref{tab:demand}. An imbalance between cyclist and vehicle demands is introduced in conflicting movements, causing cyclists and vehicles to compete for the available green time in Demand Profile I. Demand Profile II has the same demands over all movements, replicating a traffic situation where all directions are of equal importance. Another traffic situation that frequently occurs is when a high-demand corridor intersects a low-demand one. This is represented in Demand Profile III, where the east-west and west-east movements have higher demand than the other movements. 
\begin{table}[htbp]
\caption{Demand profiles given in $\frac{\mathrm{cyclist}}{\mathrm{lane}\cdot\mathrm{hour}}$ or $\frac{\mathrm{vehicle}}{\mathrm{lane}\cdot\mathrm{hour}}$.}
\begin{center}
\begin{tabular}{|c|c|c|c|c|c|c|c|}
\hline
\multirow{2}{4em}{Demand} &\multicolumn{3}{|c|}{\textbf{Demand profiles}} & \multirow{2}{4em}{Demand} &\multicolumn{3}{|c|}{\textbf{Demand profiles}} \\
\cline{2-4} 
\cline{6-8} 
  & \textbf{\textit{I}}& \textbf{\textit{II}}& \textbf{\textit{III}} &  & \textbf{\textit{I}}& \textbf{\textit{II}}& \textbf{\textit{III}}\\
\hline
 V1 & 400 & 400 & 600 & V7 & 100 & 400 & 200  \\
 V2 & 400 & 400 & 200 & V8 & 100 & 400 & 200  \\
 V3 & 400 & 400 & 200 & C1 & 100 & 800 & 800  \\
 V4 & 400 & 400 & 200 & C2 & 100 & 800 & 300  \\
 V5 & 400 & 400 & 600 & C3 & 100 & 800 & 800  \\
 V6 & 400 & 400 & 200 & C4 & 800 & 800 & 300  \\
\hline
\end{tabular}
\label{tab:demand}
\end{center}
\end{table}

Vehicle and cyclist arrivals at the intersection are generated stochastically by sampling the number of arrivals at each time step from a Poisson distribution. Each demand profile is simulated 10 times to account for the randomness in the arrival process. The duration of the simulations is 300 s, unless otherwise mentioned. The results are averaged over these 10 simulations. The nonlinear programming solver IPOPT is chosen to solve the vehicle trajectory optimisation. The parameters used in simulations are given in Table~\ref{tab:sim_parameters}.

\begin{table}[htbp]
\caption{Simulation parameters.}
\label{tab:sim_parameters}
\centering
\footnotesize
\setlength{\tabcolsep}{3pt}
\renewcommand{\arraystretch}{1.2}
\newcommand{\paramcell}[1]{\makebox[\dimexpr\columnwidth/5-2\tabcolsep\relax]{#1}}
\begin{tabular}{|*{5}{c|}}
\hline
\paramcell{$v^\mathrm{d}$ (m/s)} & \paramcell{$d_\mathrm{hw}$ (m)} & \paramcell{$Q^\mathrm{u}$} & \paramcell{$Q^\mathrm{v}$} & \paramcell{$H$ (steps)} \\
12 & 3 & 2 & 0.1 & 30 \\
\hline
$t_\mathrm{hw}$ (s) & $f_\mathrm{veh}$ (Hz) & $v_\mathrm{nor}^\mathrm{C}$ (m/s) & $u_\mathrm{des}$ (m/s$^2$) & $t_\mathrm{adj}$ (s) \\
0.5 & 1 & 5.5 & 1 & 10 \\
\hline
$D_\mathrm{SA}^\mathrm{react}$ (m) & $D_\mathrm{sig}^\mathrm{react}$ (m) & $S_\mathrm{min}$ & $S_\mathrm{max}$ & $R$ (m) \\
15 & 30 & 0.7857 & 1.8333 & 200 \\
\hline
$t_\mathrm{clr}$ (s) & $g_\mathrm{min}$ (s) & $g_\mathrm{max}$ (s) & $f_\mathrm{sig}$ (Hz) & $f_\mathrm{SA}$ (Hz) \\
3 & 6 & 25 & $1/9$ & $1/3$ \\
\hline
\end{tabular}
\end{table}

Since the communication range of 200 metres is relatively small, a prediction horizon of two traffic signal cycles was sufficient, as all road users were able to pass through the intersection within this horizon. Increasing the communication range could potentially improve performance by allowing vehicle trajectories to be optimised from farther upstream, providing more time and distance to adjust vehicle speeds to the planned green intervals. However, increasing the communication range also results in an optimisation problem containing more vehicles, thereby increasing the computational burden of the vehicle optimisation.

The performance measures used for the analysis are the Wait Time (WT), defined as the duration during which a cyclist moves slower than 1 m/s, and the Travel Time (TT), defined as the time elapsed between entering the communication range and reaching a point 50 m beyond the stop line. Measuring travel time beyond the stop line accounts for the time lost while accelerating back to normal speed.

\subsection{Impact of weight ratio and simulation time}
\label{subsec:weight}
The results of the weight ratio analysis for Demand Profile I are given in Table \ref{tab:W_ratio}, where $\mu$ and $\sigma$ denote the mean and standard deviation, respectively. The weight ratio is defined as $W^\mathrm{V}/W^\mathrm{C}$, with larger values giving greater relative priority to AVs. During this analysis, no speed advice is provided to cyclists, which is implemented by fixing the slowness factors in $\mathbf{S}$ to unity, such that cyclists travel at their normal speed. The performance of the proposed framework is compared against a fixed-time (FT) controller, which allocates green time to each phase in proportion to its traffic demand. Under this baseline, vehicle trajectories are also optimised, while the signal timings remain fixed.

Since the cyclist and vehicle movements with a high demand are conflicting in Demand Profile I, the effect of the weight ratio is most pronounced. The improvement in the mean vehicle TT increases with the weight ratio, while the improvement in the mean cyclist TT decreases. At a weight ratio of 0.5, the mean TTs of vehicles and cyclists are reduced by 13.6\% and 12.3\%, respectively, giving the most similar percentage reductions for the two modes.
\begin{table}[htbp]
\caption{Results for Demand Profile I for different weight ratios.}
\begin{center}
\begin{tabular}{|c|c|c|c|c|c|c |}
\hline
& &\multicolumn{5}{|c|}{\textbf{Weight ratio }} \\
\cline{3-7} 
 & \textbf{FT} & \textbf{\textit{1.0}} & \textbf{\textit{0.75}} & \textbf{\textit{0.5}} & \textbf{\textit{0.25}} & \textbf{\textit{0.1}}  \\
\hline
$\mu_\mathrm{WT,cyc}$ (s)    & 15.78  & 9.00  & 8.65  & 8.06  & 7.04  & 6.46  \\
$\sigma_\mathrm{WT,cyc}$ (s) & 14.48  & 9.34  & 9.14  & 8.82  & 8.30  & 8.21  \\
\hline
$\mu_\mathrm{TT,cyc}$ (s)    & 65.25  & 58.35  & 57.93  & 57.24  & 55.97  & 55.12  \\
$\sigma_\mathrm{TT,cyc}$ (s) & 16.03  & 10.90  & 10.77  & 10.50  & 10.16  & 10.18  \\
\hline
$\mu_\mathrm{TT,veh}$ (s)    & 38.43  & 32.53  & 32.73  & 33.19  & 34.06  & 35.43  \\
$\sigma_\mathrm{TT,veh}$ (s) & 15.09  & 10.26  & 10.38  & 10.59  & 11.15  & 11.97  \\ 
\hline
\end{tabular}
\label{tab:W_ratio}
\end{center}
\end{table}

\begin{figure}[htb]
    \centering
    \includegraphics[width=\linewidth]{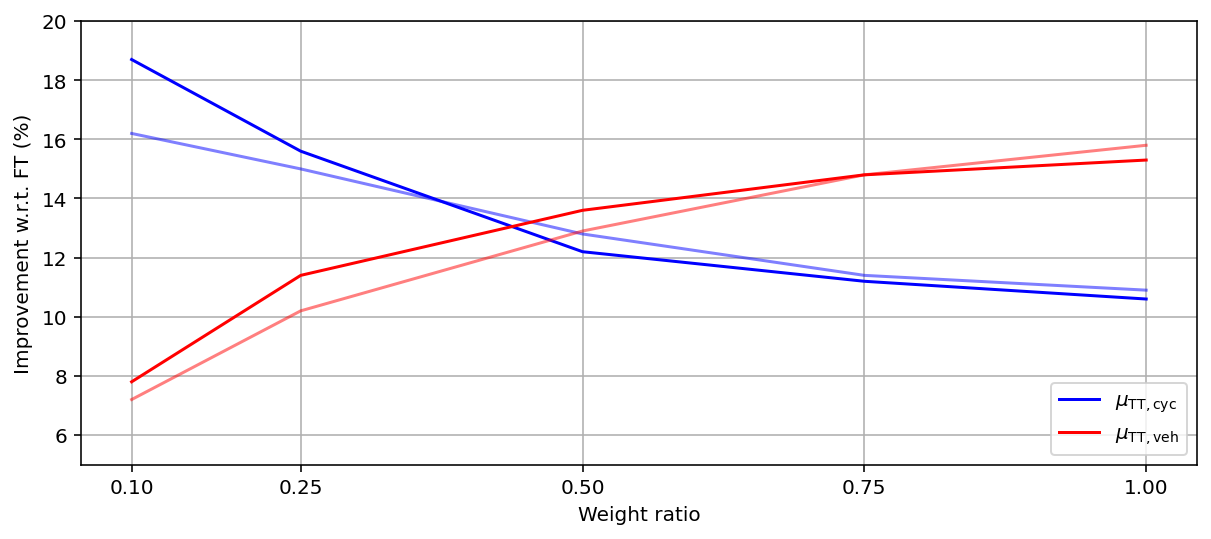}
    \caption{Mean travel time improvement with respect to FT control for different weight ratios. Solid and faded lines indicate long and short simulations, respectively.}
    \label{fig:long_sim}
\end{figure}

In addition to the 5-minute simulations, 30-minute experiments are conducted to analyse the long-term performance of the proposed method. These longer simulations use Demand Profile I and give results similar to those of the shorter simulations (see Figure \ref{fig:long_sim}).

\subsection{Impact of speed advice for cyclists}
\label{subsec:res_SA}
The speed advice is added to the hierarchical framework to reduce the waiting time of cyclists. To assess the effectiveness of the SA, the performance with different SA distances is compared with the scenario where there is no SA. The results are given in Table \ref{tab:SA_res}, where the sign position is the distance between the SA sign and the signal. 

\begin{table}[htbp]
\caption{Results of speed advice, averaged over the demand profiles.}
\begin{center}
\setlength{\tabcolsep}{4pt}
\begin{tabular}{|c|c|c|c|c |}
\hline
& &\multicolumn{3}{|c|}{\textbf{SA sign position}} \\
\cline{3-5} 
 & \textbf{No SA } &\textbf{\textit{70 m}}& \textbf{\textit{110 m}}& \textbf{\textit{150 m}}  \\
\hline
$\mu_\mathrm{WT,cyc}$ (s)    &  8.25  & 3.16 \gray{(61.7\%)} &  1.55 \gray{(81.2\%)} & 1.70 \gray{(79.4\%)}  \\
$\sigma_\mathrm{WT,cyc}$ (s) &  8.60  & 5.37                 &  4.15                 & 5.96                   \\
\hline
$\mu_\mathrm{TT,cyc}$ (s)    & 57.35  & 56.64 \gray{(1.2\%)} & 56.44 \gray{(1.6\%)}  & 56.09 \gray{(2.2\%)} \\
$\sigma_\mathrm{TT,cyc}$ (s) & 10.42  & 10.99                & 11.67                 & 13.40                  \\
\hline
$\mu_\mathrm{TT,veh}$ (s)    & 33.54  & 33.15 \gray{(1.2\%)} & 33.09 \gray{(1.3\%)}  & 33.02 \gray{(1.6\%)} \\
$\sigma_\mathrm{TT,veh}$ (s) & 10.69  & 10.44                & 10.30                 & 10.43                  \\ 
\hline
\end{tabular}
\label{tab:SA_res}
\end{center}
\end{table}

The mean WT over all three demand profiles improves substantially when the speed advice is given at 110 m or 150 m (around 80\% w.r.t. no SA), compared to when the speed advice is given at 70 m from the signal (around 60\% w.r.t. no SA). This is expected since the cyclists are close to the intersection and therefore they move at the given speed advice for a shorter duration, causing the speed advice to have less influence on the arrival time of the cyclists. The standard deviation of the TT and the WT for cyclists is larger for the 150 m case as well, compared to that for the 110 m case. This is expected, since the speed advice is given further away, and the signal timing can change due to new vehicles and cyclists entering the control area. This can lead to a higher standard deviation and higher outliers.

In the previous simulations, full compliance was assumed, meaning that all cyclists attempted to follow the speed advice. Also, zero compliance error was assumed, meaning that cyclists reached the exact advised speed. The effects of relaxing these two assumptions are evaluated in two separate experiments, as described below.

\begin{enumerate}
    \item Lowering the compliance rate: The compliance rate determines the proportion of cyclists who intentionally follow the speed advice, while the remaining cyclists ignore it completely. Complying cyclists are assumed to reach the advised speed without error.
    \item Introducing compliance error: All cyclists attempt to follow the speed advice, but their achieved speeds deviate from it by an error sampled from a normal distribution with zero mean and a standard deviation of $0.5$ m/s. This experiment assumes full compliance.
\end{enumerate}
Note that the signal optimisation continues to assume full compliance and zero compliance error in both experiments.

\begin{table}[htbp]
\caption{Results for compliance rate and error experiments.}
\begin{center}
\begin{tabular}{|c||c|c ||c|}
\hline
& \multicolumn{2}{c||}{\textbf{Compliance rate}}
& \multirow{2}{*}{\textbf{Compliance error}} \\
\cline{2-3}
& \textbf{\textit{80\%}} & \textbf{\textit{50\%}} & \\
\hline
$\mu_\mathrm{WT,cyc}$ (s)    & 3.50 & 4.58    & 2.22  \\
$\sigma_\mathrm{WT,cyc}$ (s) & 6.70 & 7.64    & 5.75     \\
\hline
$\mu_\mathrm{TT,cyc}$ (s)    & 56.88  & 57.27 & 57.18 \\
$\sigma_\mathrm{TT,cyc}$ (s) & 11.46  & 11.41 & 12.96       \\
\hline
$\mu_\mathrm{TT,veh}$ (s)   & 33.13 & 33.12   & 32.99\\
$\sigma_\mathrm{TT,veh}$ (s)& 10.44 & 10.31   &  10.43\\ 
\hline
\end{tabular}
\label{tab:SA_comp}
\end{center}
\end{table}
 
The results of the compliance rate and compliance error experiments are given in Table~\ref{tab:SA_comp}. As the compliance rate decreases, the waiting time for cyclists increases. However, when compared to the signal optimisation without speed advice (see Table~\ref{tab:SA_res}), there was a reduction of 44\% with a 50\% compliance rate. This indicates that speed advice can substantially reduce mean cyclist waiting time even under partial compliance. Introducing compliance error led to a mean waiting time of 2.2 s. However, there was an increase in the standard deviation of cyclist travel time, likely caused by a small number of cyclists having substantially longer TTs. A possible reason is that cyclists who were advised to travel at a low speed moved too slowly and missed the green light. Consequently, they had to wait until the next cycle.

\section{Conclusions}
\label{chap:conclusion}
We propose a novel hierarchical framework that integrates the optimisation of traffic signals, autonomous vehicle (AV) trajectories, and cyclist speed advice. The hierarchical framework combines the joint optimisation of signal timings and cyclist speed advice at the upper level with speed advice optimisation and AV trajectory optimisation at the lower level to improve the traffic performance of both AVs and cyclists.

We evaluate the performance of the proposed framework through simulations to examine the effects of different traffic demands, optimisation weights, speed advice sign locations, and cyclist compliance rates. The results show that signal and vehicle optimisation reduced travel times for both AVs and cyclists compared to the baseline. Adding speed advice had a limited effect on travel times but reduced cyclists' waiting time by approximately 80\% under full compliance and 44\% under a 50\% compliance rate, compared with the case without speed advice. 

Overall, the proposed framework demonstrates that cyclists can be incorporated into hierarchical traffic control without compromising the benefits for autonomous vehicles.

\begin{figure}[htb]
    \centering
    \includegraphics[width=\linewidth]{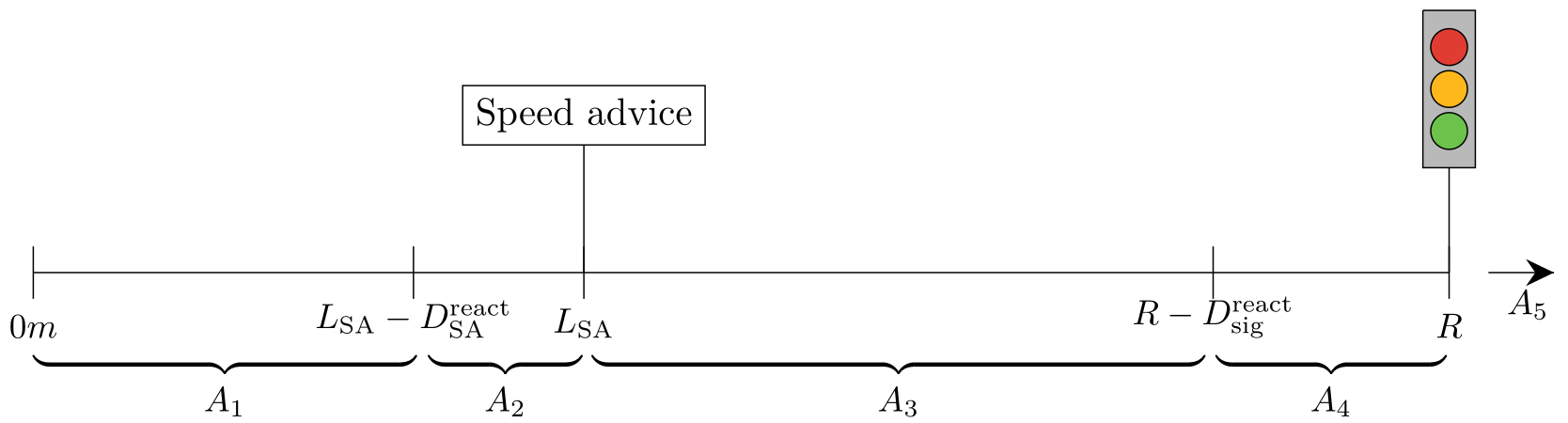}
    \caption{Spatial regions for cyclist simulation. $D_\mathrm{SA}^\mathrm{react}$ and $D_\mathrm{sig}^\mathrm{react}$ are the distances upstream of the advice sign ($L_\mathrm{SA}$) and stop line ($R$) at which cyclists start reacting, respectively.}
    \label{fig:cyc_regions}
\end{figure}

\appendix\label{chap:appendix}
Figure~\ref{fig:cyc_regions} shows the spatial regions used in the cyclist simulation. In $A_1$, cyclists cruise at their normal speed. In $A_2$, cyclists adjust to the speed advice using \eqref{eq:adj_speed}, then cruise in $A_3$. In $A_4$, cyclists decelerate to a stop according to \eqref{eq:cyc_stop} under red or when unable to cross before red during amber. Otherwise, cyclists maintain their current speed during amber. Under green, cyclists maintain their normal speed or adjust towards it if their current speed differs. In $A_5$, beyond the stop line, cyclists who stopped at a red signal accelerate back to normal speed, while cyclists who crossed the stop line at normal speed continue cruising. Adjustments to normal speed follow \eqref{eq:adj_speed} with $v_\mathrm{SA}$ replaced by $v_\mathrm{nor}^\mathrm{C}$.

\bibliographystyle{IEEEtran}
\bibliography{MyBib}

\end{document}